\documentstyle[12pt,pictex,epsfig]{article}
\newcommand{\ra}{\rightarrow}

\newcommand{\be}{\begin{equation}}
\newcommand{\ee}{\end{equation}}
\newcommand{\bea}{\begin{eqnarray}}
\newcommand{\eea}{\end{eqnarray}}

\newcommand{\ov}{\overline}
\def\npb#1#2#3{    {\it Nucl. Phys. }{\bf B #1} (19#2) #3}
\def\plb#1#2#3{    {\it Phys. Lett. }{\bf B #1} (19#2) #3}
\def\prd#1#2#3{    {\it Phys. Rev. }{\bf D #1} (19#2) #3}

\def\ijmpa#1#2#3{  {\it Int. J. Mod. Phys. }{\bf A #1} (19#2) #3}

\begin{document}
\thispagestyle{empty}   
\begin{flushleft}
Oslo TP 8-95
\end{flushleft}
\vspace{0.4cm}

\vspace{0.5cm}

\begin{center}
  \begin{Large}
  \begin{bf}
Momentum Dependence of the Penguin Interaction
  \end{bf}
  \end{Large}
\end{center}

  \vspace{0.5cm}
  \begin{Large}
\begin{center}     
A. E. Bergan and J. O. Eeg
\end{center}
\end{Large}

\begin{Large}
\begin{center}
{\small  Department of Physics, University of Oslo, \\

P.O.Box 1048, N-0316 Oslo, Norway \\}
\end{center}
\end{Large}
\vspace{0.6cm}
\begin{center}
  {\bf Abstract}
\end{center}
\begin{quotation}
\noindent
We have considered the  penguin   interaction contribution to
 $K \rightarrow 2 \pi$ decays. In particular, we have investigated
 the effect of   the momentum dependence of the
  penguin coefficient. Our
analysis is performed within the Chiral Quark Model
 where 
quarks are coupled to the
pseudoscalar mesons, which 
 means that hadronic matrix elements can be calculated in
terms of quark loop diagrams.

We have inserted the momentum dependent penguin coefficient into
the relevant  quark loop 
diagrams for $K \rightarrow \pi$.
 We discuss
two possible prescriptions for performing the calculations, and  conclude 
 that  the momentum
 dependence of the penguin coefficient increases the amplitude 
by 10-20 $\%$.
In any case, the  (CP-conserving) penguin  contribution is very sensitive
to the values chosen for the involved parameters.
 
\end{quotation}

\newpage

\begin{Large}

1. Introduction

\end{Large}

\vspace{0.5cm}

The penguin (gluonic monopole) diagram was proposed as an
explanation of  the $\Delta I =1/2 $ rule \cite{SVZPa}. 
This
diagram induced pure $\Delta I =1/2 $ four quark operators into the
effective Lagrangian relevant for $K \ra 2\pi$ decays and other
$\Delta S = 1$ transitions. In general,
non-leptonic decays are  described by an effective 
Lagrangian\cite{NLept}
\begin{equation}
{\cal{L}}(\Delta S =1)  = - \frac{1}{\sqrt{2}} G_F \lambda_u 
\sum_{i} C_i Q_i  \; \;  ,
\label{efflag}
\end{equation}
where $G_F$ is Fermis coupling constant and $\lambda_u = V_{su} V^*_{du}$
is the relevant
Kobayashi-Maskawa\cite{KM} factor for CP-conserving
$\Delta S = 1$ transitions.
 In such effective Lagrangians, the heavy
mass scales are integrated out and their effects are contained in the Wilson
coefficients $C_i$. The  quark operators $Q_i$ involve the
three light quarks $q = u,d,s$. The coefficients will in general
include short distance QCD effects calculated by means of
perturbation theory and the
renormalization group equations (RGE). Both the coefficients $C_i$ and the
matrix elements of the quark operators depend on the
renormalization scale $\mu$ (taken to be of order 1 GeV),
in such a way that the physical processes are independent of $\mu$.
Concerning the discussion on the importance of the penguin
interaction  for the $\Delta I = 1/2$ rule within the standard approach,
 we refer to the 
literature\cite{NLept,DonOr,BBG,cheng,joe,galic,PiJa,joer}. In this
paper, we will revisit this issue from some more unconventional
point of view, and ask if some effects have been left out.

\vspace{0.2cm}

To calculate the matrix elements of quark operators between physical
had\-ronic states is in general a difficult  task, and one  normally
 uses  various models or assumptions.
In this paper we will
 use the Chiral Quark Model ($\chi QM$),  advocated by many
authors \cite{WeMaBi,ERT,PdeR}.
The model consists of the ordinary QCD Lagrangian and 
in addition a term ${\cal{L}}_{\chi}$. 
 This new term
 includes the Goldstone meson octet in  a
chiral invariant way, and provides   meson-quark couplings
that makes it possible to calculate matrix elements of quark
operators in terms of quark loop diagrams.
The model is thought to be applicable for quark momenta
 below some scale of the order $\Lambda_{\chi}$,  the  chiral symmetry
 breaking scale. Although this
is a model, we find it very interesting because it reproduces the
 chiral Lagrangian terms for strong interactions to good  accuracy to order
$p^4$ \cite{ERT}.
 As a field-theoretical model, it can be formulated in terms
of path integrals. Integrating out the quarks, one obtains chiral
Lagrangians. Alternatively, integrating out the Goldstone mesons, one obtains 
Nambu type  models\cite{Ch/NJL,BBdR}.

\vspace{0.2cm}

In order to predict physical amplitudes like $K\ra 2 \pi$, one
has to match the high energy description, contained in the
coefficients $C_i$,   with the low energy description. This
matching should be done at some scale $\mu$ where both
descriptions are valid. How to do this has been a delicate question,
and especially for the penguin contributions because the coefficients
are very sensitive to the quantity $m_c/\mu$, where $m_c$ is the
charm-quark mass. 

\vspace{0.2cm}

In this paper we will focus on the following aspect of the effective
Lagrangian (\ref{efflag}):
  The coefficients $C_i$ in (\ref{efflag}) originates from some Feynman
diagrams and will in principle depend on the external quark momenta, that is,
the quark momenta corresponding to the quark fields in $Q_i$ in
(\ref{efflag}). 
Working within the $\chi QM$,
we might keep the full momentum dependence of the penguin coefficient,
calculated to lowest order in the strong coupling constant.
 Being a field theoretic model,
 the $\chi QM$ can of course account for such
momentum dependent effects.
Unfortunately, our calculation will have some limitations. 
Especially, we lose the
 RGE analysis of the Wilson coefficient because the RGE analysis does not
 keep track of the individual quark momenta, only their overall scale.
Thus, our procedure will  be less systematic
 than the standard
one \cite{NLept,DonOr,BBG,cheng,PdeR}. 
Still, because the standard method might have some difficulties with the
matching around 1 GeV,
 we find it fruitful to perform such a calculation to
see if there are effects which are lost in the standard approach.

\vspace{1cm}

\begin{Large}

2. The penguin interaction  at quark level

\end{Large}

\vspace{0.2cm}

The  penguin diagram is shown in Fig.1, and induces
 an effective interaction (in the CP-conserving case)
\begin{equation}
{\cal{L}}_{P} \, = \, -  \sqrt{2} \, G_F \, \lambda_u \; C_P \; Q_P \;
\label{lapeng} 
\end{equation}
where $C_P$ is the Wilson coefficient of the operator
\begin{equation}
Q_P \, = \,     
 (\bar{d} \gamma_{\nu} L t^a s) \,
 \bar{q} \gamma^{\nu} t^a q \: ,
\label{openg}
\end{equation}
where a sum over $q=u,d,s$ is understood. The $t^a$'s are the colour
 matrices (for $a=1,..,8$), 
and $g_s$ is the strong coupling.
 For the CP-conserving case considered here, the dominating part is
due to  u- and c-quarks
 running in the penguin loop,  and the following expression is obtained
 in the leading logarithmic approximation before any RGE analysis
is performed:
\begin{equation}
C_P \; =  \;  - \frac{\alpha_s}{3 \pi} \, \ln \frac{m_c^2}{\mu^2}  
\label{llpeng}
\end{equation}
where $m_c$ is the c-quark mass.
Going  beyond the leading logarithmic approximation, one finds
the following  expression \cite{var,joer}: 
\begin{equation}
C_P \; =  \;  - \frac{\alpha_s}{3 \pi} \, ( \ln \frac{m_c^2}{\mu^2} +
\frac{5}{3}) \; .
\label{nlpeng}
\end{equation}
where $\alpha_s = g_s^2/4\pi$. Here the constant next to leading term 
is numerically as
important as the leading logarithm. This illustrates the 
difficulties in treating the penguin diagram perturbatively.

The expression (\ref{nlpeng}) still does not take into 
account the variation of $C_P$ with the gluon momentum $p$.
This variation can be expressed as \cite{SeP}:
\be
C_P(p^2) =  (-\frac{\alpha_s}{3\pi}) \, 6 \int^1_0 dt \, t (1-t)
 \int^{m_c^2}_{m_u^2}  \frac{d \rho}{\rho - t(1-t)p^2} \; ,
\label{fpeng}
\ee
which exhibits the GIM mechanism \cite{GIM} explicitly.
The variation of this function with momentum is shown in Fig.2.
Eq. (\ref{nlpeng}) is obtained from (\ref{fpeng}) if one uses the 
identification $p^2 = - \mu^2$, and the approximation
$m_c^2 \gg \mu^2 \gg m_u^2$.
For momenta of order $M_W$, the coefficient $C_P$ is more 
complicated\cite{joer}, but this is irrelevant for us here.
Using dimensional regularization $C_P(p^2)$ takes the following form: 
\be
C_P(p^2) =  (-\frac{\alpha_s}{3\pi}) \, 6 \, \Gamma(\epsilon)
\,  \tau(\epsilon) \, \int^1_0  dt \, t (1-t) \,
 [(-D_u)^{- \epsilon} - (-D_c)^{- \epsilon}] \; , 
\label{drpeng}
\ee
where $\Gamma$ is the gamma-function satisfying $\Gamma(n) = (n-1)!$
for an integer $n$. For space dimension $D = 4 - 2 \epsilon$, one has 
$\Gamma(\epsilon)= 1/\epsilon + constant$. Moreover, $\tau(\epsilon)
= (-4\pi \mu^2)^\epsilon$, where $\mu$ is a parameter with dimension
mass which enters the calculation within the dimensional
 regularization scheme. This parameter
 may be identified with the renormalization scale. The quark masses
 enters the expression (\ref{drpeng}) through the quantity
\be
D_q = m_q^2 - t(1-t) p^2  \; ,
\ee
for $q = u,c$.

 Using the SU(3) matrix relation
\be
t^a_{ij} t^a_{kl} \; = \; \frac{1}{2} (\delta_{il} \delta_{jk}  
\; - \; \frac{1}{N_c} \delta_{ij} \delta_{kl}) \; ,
\ee
splitting the quark current in (\ref{openg}) in a left- and right-handed part
and using a Fierz transformation, one obtains four different contributions to
${\cal{L}}(\Delta S =1)$. The most important penguin four quark operator is
 $Q_6$, which in the Fierz-transformed version is a product of two quark
densities, involving both left- and right-handed quark fields:
\be
Q_6 \; = \; -8 \, \sum_q (\bar{d_L} q_R) \, (\bar{q_R} s_L) \; ,
\label{Q6peng}
\ee
where the sum runs over $q =u,d,s$. In the naive leading
 logarithmic approximation (without
RGE), we can do the following identification.
\be
C_6 \; = \; \frac{1}{4} \, C_P  \;
\label{C6peng}
\ee
The expression for $C_6$ including RGE can be found in the literature
\cite{NLept,BBG,PdeR,BuCi}.  There are also penguin
operators which are products of two left-handed currents, like
\be
Q_4 \; = \; 4 \, \sum_q (\bar{d_L} \gamma^\mu q_L) \, 
(\bar{q_L} \gamma_\mu s_L) \; ,
\label{Q4peng}
\ee
but these are not so important numerically.

In addition to penguin operators, there is
an important  pure $\Delta I = 1/2$ operator of left-left
type which is $Q_- = Q_2 - Q_1$, where
\be
Q_1 \; = \; 4 \, (\bar{d_L} \gamma^\mu  s_L) \, (\bar{u_L} \gamma _\mu u_L)
 \; \; \mbox{and} \; \;
Q_2 \; = \; 4 \, (\bar{d_L} \gamma^\mu  u_L) \, (\bar{u_L} \gamma _\mu  s_L)
\label{QLL}
\ee
are quark operators contained in (\ref{efflag}).

\vspace{0.6cm}

\begin{Large}

3. The Chiral Quark Model

\end{Large}

\vspace{0.2cm}

In the Chiral Quark Model ($\chi QM$) \cite{WeMaBi,ERT,PdeR},
 chiral-symmetry breaking is thought to be
taken into account by adding an extra term 
${\cal{L}}_{\chi}$ to ordinary QCD:
\begin{equation}
{\cal{L}}_{QCD} \rightarrow {\cal{L}}_{QCD\chi} = {\cal{L}}_{QCD}
+  {\cal{L}}_{\chi} \; \; \; ; \; \; \;  {\cal{L}}_{QCD} = {\cal{L}}_{f}
+  {\cal{L}}_G\ ,
\label{chQM}
\end{equation}
where
\begin{eqnarray}
 {\cal{L}}_{f} = \bar{q} (i \gamma \cdot D - {\cal{M}}_q) q \; \; \;
; \; \; \; \; \;
q  \; = \;
 \left( \begin{array}{c} u \\  d\\  s \end{array}
\right)  \; \; , \;
\label{Lf}
\end{eqnarray}
 ${\cal{M}}_q$ is the {\em current} quark mass-matrix,
${\cal{L}}_G$ is the pure gluonic part of QCD, and 
\begin{equation}
  {\cal{L}}_{\chi} = - M \, (\overline{q_L} \Sigma q_R + 
\overline{q_R} \Sigma^\dagger q_L) \; .
\label{Lchi}
\end{equation}
 The constant  $M$ in (\ref{Lchi}) is interpreted as the
{\em constituent} quark mass, thought to be of order 200 -  300 MeV.
Note that the constituent and current masses are connected to
different terms in the Lagrangian, with different transformation properties.
The quantity $\Sigma$ contains
 the Goldstone-octet fields $\pi^a$:
\begin{equation}
\Sigma = exp(i\sum_a \lambda^a \pi^a / f) \; \; ,
\label{sig}
\end{equation}
where $\lambda^a$
are the Gell-Mann matrices, and $f \, = \, f_{\pi} = 93$ MeV is the pion
decay constant.
 The term ${\cal{L}}_{\chi}$ introduces  meson-quark couplings.
This means that the quarks can be integrated out and the coefficients
of the various terms in the effective meson theory, 
  the chiral Lagrangian, are calculable
 from ${\cal{L}}_{QCD\chi}$ \cite{ERT}. It has  also been found
to be suitable for calculating
hadronic matrix elements of operators obtained from the weak
 sector\cite{PdeR,BEF}, like in eq. (\ref{efflag}).

There are several versions of the $\chi QM$ in the literature.
 Note that in eqs. 
(\ref{chQM})-(\ref{sig}), there is no kinetic term for the mesons.
Thus, the meson fields are external. They propagate only after the quarks are
integrated out. 
The $\chi QM$ is thought to
apply  for  momenta of the order and below  
the scale of chiral-symmetry breaking, which we define to be
\begin{equation} 
\Lambda_{\chi} = 2 \pi f_\pi \sqrt{6/N_c} \; ,
\label{lambdachi}
\end{equation}
 where $N_c = 3$ is
the number of colours (-numerically $\Lambda_{\chi}$ = 0.83 GeV).
Therefore a physical ultraviolet cut-off   $\Lambda$ of the  order
 $\Lambda_{\chi}$ is sometimes used within the $\chi Q M$ to parametrize
loop integrals which would otherwise be divergent. Such divergences
will be buried in the physical $f_\pi$ and quark condensate (as shown below),
 and are not  removed by counterterms as in ordinary field theories.
One should note that   $\Lambda$ and $\Lambda_{\chi}$ are in principle
different quantities.
One may also use dimensional regularization within the $\chi QM$, as
shown in \cite{BEF}.
Even if dimensional regularization is used,
$\Lambda_{\chi}$ is still some kind of  effective cut-off scale
 of the $\chi QM$, just as for chiral perturbation theory.
(Analogously, the W-mass is some effective cut-off scale of electroweak
interactions, even if dimensional regularization is used).
 
 There are a priori several ways to introduce an
  ultraviolet cut-off $\Lambda$ within the $\chi Q M$.
The simplest way is to use an explicit
(sharp) cut-off in the integration over virtual Euclidean momenta.
However, this prescription
 violates translation invariance, and will only
give a unique result for the leading term. One might try
 a Pauli-Villars type of cut-off\cite{Pau-Vi}, which turns out
to be technically cumbersome in our case.
 A good alternative is proper time regularization, which has already been
 used \cite{BidR,BBdR} in $\chi QM$ 
calculations. In this case one uses the
following replacement for the propagator denominator (for Euclidean momenta):
\begin{equation}
\frac{1}{p^2 + M^2}  \; \rightarrow \; 
\int^{\infty}_{\xi} d \tau \, exp[- \tau (p^2 + M^2)] \; ,
\label{prtim}
\end{equation}
where $\xi =  1/\Lambda^2$.

The cut-off $\Lambda$ can not be chosen freely. Indeed
the $\chi QM$\cite{PdeR,BBdR} provides a relation between 
$\Lambda$, $M$ and $f_{\pi}$, and eventually gluon condensates.
Such a relation is obtained because $f_{\pi}$, entering the meson-
quark coupling $\sim M \gamma_5/f_{\pi}$, is also given by a
quark loop diagram for $\pi \ra W$(virtual), and one obtains:
\begin{equation}
f^{(0)}_{\pi} \, = \, \frac{N_c M^2}{4 \pi^2 f} 
[ \; \hat{f}_{\pi}
\;  + \frac{\pi^2}{6 N_c M^4}
<\frac{\alpha_s}{\pi} G^2> + \cdots ] \; ,
\label{fpi}
\end{equation}
where  $<\frac{\alpha_s}{\pi} G^2>$
 is the two gluon condensate, and the dots indicate higher gluon
condensates. The value of $\Lambda$ will depend on how many gluon
condensates which are kept in (\ref{fpi}).
In the end both $f$ and $f^{(0)}_{\pi}$ will, in the
limit $m_{u,d} \ra 0$, be identified by $f_{\pi}$, but at
intermediate stages one might need to distinguish them for
technical reasons (-we also have $f_K = f^{(0)}_{\pi}$ in the limit
$m_{u,d,s} \rightarrow 0$).

The dimensionless quantity $\hat{f}_{\pi}$ has the leading behaviour 
$\sim \ln(\frac{\Lambda^2}{M^2})$ in a cut-off type
regularization. Its explicit form will depend on the
way the cut-off is introduced. Within dimensional regularization,
$\hat{f}_{\pi} = \Gamma(\epsilon) (4 \pi \tilde{\mu}^2/M^2)^\epsilon$,
where $\tilde{\mu}$ is a parameter with dimension mass, occurring
owing to the use of dimensional regularization in the $\chi QM$.
Thus, the would be divergent (for $\Lambda \rightarrow \infty$)
logarithmic term (or $1/\epsilon$ term) is absorbed in the physical
pion decay constant $f_{\pi}$ \cite{PdeR,BEF}.

The quark condensate is within $\chi QM$ given by\cite{PdeR,BBdR,Pau-Vi,RR}
\begin{equation}
<\bar{q} q> \, = \, - \, (\tilde{\mu}^2)^{\epsilon} \, 
\int \frac{d^D p}{(2 \pi)^D} 
Tr[i S(p)]  \: = \:  \frac{N_c M}{4 \pi^2} C_q
 - \frac{1}{12 M} <\frac{\alpha_s}{\pi} G^2> + ...  \ ,
\label{qcond}
\end{equation}
where $S(p)$ is the quark propagator in an external gluon field\cite{RR}.
 The  quantity
$C_q$ depends on  the regularization prescription.
For a  cut-off type regularization, $C_q = - \Lambda^2  + 
M^2 \ln \frac{\Lambda^2}{M^2} + \cdots$
Within dimensional regularization, 
$C_q = - M^2 \Gamma(-1+\epsilon) (4 \pi \tilde{\mu}^2/M^2)^\epsilon$.
Equation (\ref{qcond}) puts (in addition to eq.(\ref{fpi})) further
restrictions on the parameters of the $\chi QM$.


\vspace{1cm}

\begin{Large}

4. The standard $K \rightarrow \pi$ amplitude. 

\end{Large}

\vspace{0.2cm}

The $K \rightarrow 2\pi$ amplitudes can in general be described
in terms of chiral Lagrangians. Knowing the structure of chiral Lagrangians,
the octet part of the $K \rightarrow 2\pi$ amplitude
 can be found from the virtual
$K \rightarrow \pi$ amplitude \cite{DonOr,BBG,cheng,PdeR}.
Following \cite{PdeR} we define the coupling constant $g_8^{(1/2)}$
of the octet chiral Lagrangian. In term of this, the virtual
$K \ra \pi$ amplitude can be written:
\be
 {\cal{M}}(K^-  \rightarrow \pi^-)_{8} \; = \;  - \, \sqrt{2} G_F
\lambda_u \, k^2 \, g^{(1/2)}_{8} \, f_K \, f_{\pi} \; ,
\label{amp}
\ee  
where $k$ is the momentum of the virtual meson transition.

The standard result for the penguin contribution, mainly due to $Q_6$,
 may be written as \cite{PdeR}
\be
g(Q_6)_S \equiv g^{(1/2)}_{8}(Q_6)_S = - 16 \, Re C_6(\mu)  
     \;  (\frac{\langle \ov{q} q \rangle}{f^3})^2 \, L_5 \; ,
\label{gQ6PR}
\ee
where $L_5$ is the coupling constant of the relevant term in the
strong chiral Lagrangian of ${\cal{O}}(p^4)$. Within the $\chi QM$
one obtains \cite{BBdR,BEF}
\be
 L_5 \; = \; - \,
  \frac{f^3 f_\pi^{(0)}}{8 M \langle \ov{q} q \rangle} 
\, \left[ 1- \rho \right] \; .
\label{cL5}
\ee
where the quantity $\rho$ is given 
(up to ${\cal{O}}(M^4/\Lambda_{\chi}^4)$) by
\be
 \rho \; = \;
   6 \frac{M^2}{\Lambda_{\chi}^2} \frac{f}{f_\pi^{(0)}}  
  \; ; \qquad 
\mbox{or} \qquad \rho \; = \; \frac{1}{\Gamma(0,x)} \; ,
\label{rho}
\ee
where the first  expression is used within dimensional regularization
in \cite{BEF}, and the last one
involving  the incomplete $\Gamma$-function,  $\Gamma(0,x)$
with $x= M^2/\Lambda^2$, is used
within proper time regularization in \cite{BBdR}. In fact the two
expressions for $\rho$ are equivalent when using eq. (\ref{lambdachi}),
and (\ref{fpi}) with $\hat{f_\pi} = \Gamma(0,x)$. Keeping 
$f_\pi^{(0)}/f = 1$ in the first expression for $\rho$ in (\ref{rho}),
the expression  for $L_5$ in (\ref{cL5}) is very sensitive to $M$.

Within the $\chi QM$, the quantity $g^{(1/2)}_{8}(Q_6)$ is given by the
diagrams in Fig.3. The $K s \bar{u}$ and $\pi \bar{u} d$ vertices are
obtained  from ${\cal{L}}_{\chi}$ in (\ref{Lchi})-(\ref{sig}).
These diagrams  contain  terms
independent of the meson momentum. However, such terms cancel in
accordance with chiral symmetry \cite{DonOr} .
 The leading physical amplitude corresponds
to the sum of the terms of order $k^2$.
The leading term  in the expression for $L_5$ in (\ref{cL5})
corresponds to diagram 3a, while the formally non-leading term
$\sim \rho$ in (\ref{cL5}) corresponds to diagram 3b. Numerically,
however, there is a rather strong cancellation between these two terms
for  values of $M$ bigger than 250 MeV.
We observe that values of $M \simeq$ 330 MeV or bigger
are not acceptable (while keeping $f_\pi^{(0)}/f = 1$)
because they correspond to zero or even slightly
negative values of $L_5$ in (\ref{cL5}).
Using explicitly the second expression for $\rho$ in (\ref{rho})
would lead to a reduced value of $L_5$ for $M$ smaller than 250 MeV.
One should however take into account that $\Gamma(0,x) = \hat{f_\pi}$
is related to the physical $f_\pi$ through (\ref{fpi}).
It is a priori an
open question if the strong cancellation between the leading 
diagram 3a and the nonleading 3b
 persists when the momentum
variation of the penguin coefficient is taken into account 
(- diagram 3c serves to cancel the momentum independent parts of 3a and 3b).

In \cite{PdeR} the expression (\ref{cL5}) was not used. Instead  the value
 $L_5= 1.8 \times 10^{-3}$ was extracted from other arguments.
Moreover, using $\mu \simeq M$, and the scale independent quark condensate
$<\bar{q} q> = (-194 \, {\mbox MeV})^3$,  it is found that
$g(Q_6)_S = 0.26$.
In general,  one obtains bigger values for  the quark condensate 
if one uses the expression  in \cite{BBG},
\be
\langle \ov{q} q \rangle (\mu) \; = \; - \, 
\frac{f_\pi^2 m_\pi^2}{m_u(\mu) + m_d(\mu)} \; \; .
\label{BBGcond}
\ee
Alternatively, using the values\cite{deRaf} $<\ov{q} q> = (-235{\mbox MeV})^3$,
$L_5 = 1.4 \times 10^{-3}$ at the scale 1 GeV (- corresponding to
 $M \simeq$ 245 MeV if (\ref{cL5}) is used),
and using \cite{BuCi}
 $C_6(\mu = 1 {\mbox GeV})
 = -0.026$ for $\Lambda_{QCD}$= 400 MeV, one obtains $g(Q_6)_S = 0.15$.
However, bigger values might be obtained.
For $\mu \simeq$ 0.8 GeV and $\Lambda_{QCD} =$ 350 MeV, 
$C_6 \simeq -0.1$, and using in addition \cite{BEF,Trst}
  $M \simeq$ 200 MeV, one obtains $g(Q_6)_S \simeq$ 0.8. 
(Note that, from (\ref{nlpeng}) and (\ref{C6peng}) one obtains
 $C_6 \simeq -0.05$ for $\mu \simeq \Lambda_\chi$)
 
 The biggest contribution to $g^{(1/2)}_{8}$  
is coming from the four quark operator $Q_{-}$ of left-left type
(-see (\ref{QLL})), which gives the amplitude\cite{PdeR}
\be
g^{(1/2)}_{8}(Q_{-}) \; = \; \frac{1}{2} C_- [1 - \frac{1}{N_c}(1 - \delta)]
\label{gG2}
\ee
where $C_-$ is the Wilson coefficient of the operator $Q_-$, which is
$\simeq 2$ at $\mu \simeq $ 0.8 GeV, and the quantity $\delta$ represents
 non-factorizable gluon condensate corrections:
\be
\delta \, \equiv \,  
\frac{N_c <\frac{\alpha_s}{\pi} G^2>}{32 \pi^2 f_\pi^4} \; .
\label{delta}
\ee
Writing $<\frac{\alpha_s}{\pi} G^2> =  \eta^4$, the quantity 
$\eta$ is of the order 400 MeV, which gives a value of $\delta$
around 3 ($\delta \simeq 2.6 \, , 3.0 \,$ , and 3.7, for
 $\eta = 376 \, , 390 \, ,$ and 410 MeV, respectively.)
Using $\eta =$ 376 MeV and $\mu \simeq $ 0.8 GeV
one obtains  $g^{(1/2)}_{8}(Q_{-}) \simeq 1.6$, while for
$\eta =$ 390 MeV and $\mu \simeq $ 320 MeV one obtains\cite{PdeR}
 $g^{(1/2)}_{8}(Q_{-}) \simeq 2.6$.  Anyway, the prediction will be below 
the experimental value  for the total $\Delta I = 1/2$ amplitude 
\cite{BBG,PdeR}  
\be
g^{(1/2)}(Tot.)_{Exp.} \; = \; 5.1 \; \;  .
\label{gex}
\ee


\vspace{1cm}

\begin{Large}

5. Momentum dependent penguin coefficient

\end{Large}

\vspace{0.2cm}

Now we will present the calculation using a momentum dependent 
penguin coefficient.
In the chiral limit $m_{s,d} \rightarrow 0$,
we obtain the following amplitude 
corresponding to the operator $Q_6$ from the diagram in Fig.3a,
\begin{equation}
{\cal{M}}_a(K^-  \rightarrow \pi^- ; Q_6) \; = \; \sqrt{2} G_F
\lambda_u \, (-8) \,
(\frac{N_c \, M}{f_{\pi}} \, \tilde{\mu}^{2\epsilon})^2 
  \, \, I_{\chi} \; ,
\label{amp6} 
\end{equation}
 where $I_{\chi}$ is the two loop integral 
\begin{equation}
I_{\chi} \; = \;  
\int \frac{d^D p}{(2\pi)^D}\int \frac{d^D r}{(2\pi)^D}
\, F(s',s) \, [C_P(p^2)] \, F(r',r) \; .
\label{chint}
 \end{equation}
Here $s$ and $r$ are loop momenta and $k = r-r' = s-s'$ is the
momentum of the virtual $K \ra \pi$  transition,
and $p = s-r$ is the momentum of the penguin gluon. 
 The function $F$ is given by
\begin{equation}
F(r',r) \, = \, \frac{r \cdot r' - M^2}{(r^2-M^2)(r'^2-M^2)}
\label{ffunc}
 \end{equation}
where the trace in Dirac space has been taken within naive 
dimensional regularization, for simplicity. 
If $C_P$ is taken as a constant, $I_{\chi}$ splits in the product
of two one loop integrals, and we recover the result of other 
authors\cite{NLept,DonOr,PdeR,BEF}. Taking $C_P$ as momentum
dependent, $I_{\chi}$ is a two loop integral and cut-off
procedures are in general cumbersome to implement.
We have, however, found that
if $C_P$ is given as in (\ref{drpeng}), the integral can be calculated
analytically within dimensional regularization (in $D = 4-2 \epsilon$
dimensions). Then the contribution from the diagram in Fig.3a
given by (\ref{amp6}),(\ref{chint}) and (\ref{ffunc})  turns out to
  dominate over  the contribution from the diagram in Fig.3b.  

We have calculated the integral using Feynman parametrization in
the usual way. The $r$ integration is easily performed, while the
remaining integration over $p$ is the tricky part. Some details
will be given in the Appendix.
Within the approximation $m_c^2 \gg M^2$, we find the result: 
\be
 I_{\chi} \, = \, (- \frac{\alpha_s}{3\pi})  (\frac{i}{16 \pi^2})^2
\; k^2 \, (-2) \,(\frac{1}{\epsilon})^2 \, m_c^2
\,  [1 + {\cal{O}} (\frac{M^2}{m_c^2}) + {\cal{O}}(\epsilon)] \; ,
\label{chint2}
\ee 
Here we have dropped an irrelevant term constant with respect to
the virtual meson momentum $k$. This constant cancels with the
constant term obtained from the diagram in Fig.3b and 3c. We find
that the  $\sim k^2$
contribution from  diagram 3b is  $\sim M^2$ instead of $\sim
m_c^2$. 
Examining the integral $I_{\chi}$ to order $k^2$, one
observes (-see Appendix) that
one factor $1/\epsilon$ comes from a quadratic divergence and
will be related to the quark condensate, and the other
$1/\epsilon$ comes from a logarithmic divergence and will be
related to $f_\pi$. Now we can combine the result (\ref{chint2})
with the equations (\ref{fpi}) 
and  ( \ref{qcond}), and we obtain a result for 
${\cal{M}}_a(K^-  \rightarrow \pi^-; Q_6)$ in (\ref{amp6}).
 Then we can convert this result  into the following
contribution to  $g^{(1/2)}_8$:
\be
g(Q_6)_{DR} = - \frac{\alpha_s}{3 \pi} 
\frac{\langle \ov{q} q\rangle m_c^2}{M^3 f_{\pi}^2} 
[1 + {\cal{O}}(\frac{M^2}{\Lambda_\chi^2}) + {\cal{O}}(\frac{M^2}{m_c^2})]
  \; ,
\label{gQ6BE}
\ee 
where the ${\cal{O}}(M^2/\Lambda_\chi^2)$ term are coming from
 ${\cal{O}}(\epsilon)$ in (\ref{chint2}).
 The result in (\ref{gQ6BE}) has to be
 compared to the standard amplitude
\be
g(Q_6)_{S} =  - \frac{\alpha_s}{3 \pi} \,
 ( \ln \frac{m_c^2}{\Lambda_{\chi}^2} + \frac{5}{3}) \,
\frac{\langle \ov{q} q \rangle}{2 M f_\pi^2} 
\, \left[ 1- 6 \frac{M^2}{\Lambda_{\chi}^2}\right] \; ,
\label{gQ6S}
\ee  
obtained from combining (\ref{nlpeng}),(\ref{gQ6PR})
and (\ref{cL5}). It might be surprising 
 that the amplitude (\ref{gQ6BE}) is found to be proportional to $m_c^2$,
 which means a huge
enhancement compared to (\ref{gQ6S}) from the leading term alone.
 This behaviour
reflects that loop momenta of order $m_c$ are important in
 the last loop integral over $p$.
The result  is very sensitive to the involved parameters.
Using $\alpha_s = \alpha_s(m_c^2) \simeq 0.38$ , 
$\langle \ov{q} q\rangle = (-257MeV)^3$ at $\mu = m_c$, we obtain
$g^{(1/2)}_{8} (Q_6) \simeq 10$ for $M$= 0.25 GeV and 
$g(Q_6) \simeq 19$ for $M$= 0.2 GeV, which is far above
 the experimental value for $g^{(1/2)}$!
However, we observe from our calculations that the formally non-leading terms 
${\cal{O}}(M^2/\Lambda_\chi^2)$ in
 (\ref{gQ6BE}), being for instance of the form
\be
\frac{M^2}{\Lambda_\chi^2} \ln \frac{m_c^2}{\Lambda_\chi^2}  \; \; ,
\qquad  \qquad
\frac{M^3 f_\pi^2}{\Lambda_\chi^2 <\bar{q} q>} 
\ln \frac{m_c^2}{\Lambda_\chi^2}  \; \; ,
\qquad \mbox{and} \qquad
\frac{M^2}{\Lambda_\chi^2} \ln \frac{\Lambda_\chi^2}{M^2}  
\label{Nonl}
\ee
might have  sizeable coefficients of order 10.
 Obviously, big cancellations 
between  leading and formally nonleading terms in (\ref{gQ6BE}) has to
occur in order to get a realistic result.  For momentum independent Wilson
 coefficients, dimensional
 regularization works pretty well \cite{BEF}.
 Unfortunately, the coefficients
(-and their signs!) of the formally nonleading terms 
in (\ref{Nonl}) have some ambiguity because
dimensional regularization does not distinguish clearly between
logarithmic divergences (associated with $f_\pi$) and quadratic
divergences (associated with the quark condensate).
Because of this, the result (\ref{gQ6BE}) is out of control and cannot be used
for predicting $g(Q_6)$. As a curiosity, we observe that
 while the standard
 amplitude $g(Q_6)_S$ has a leading term proportional to
 $\Gamma(-1+\epsilon) \cdot \Gamma(\epsilon)$, the leading term in
 $g(Q_6)_{DR}$, being the result of a two loop integration,
 is proportional to  
$\Gamma(-1+ 3 \epsilon) \cdot \Gamma(- \epsilon) \cdot F(\epsilon)$,
where $F(\epsilon)$ is finite in the limit $\epsilon \rightarrow 0$.

\vspace{0.2cm} 
The quadratic divergence is only appearing in left-right operators
like $Q_6$ in (\ref{Q6peng}). From penguin operators containing two
left-handed currents, one obtains a product of two logarithmic divergences,
and there is no  ambiguity as for quadratic divergences. The standard
contribution from $Q_4$ is (compare with (\ref{gQ6S}))  
\be
g(Q_4)_{S} =   \frac{\alpha_s}{12 \pi} \,
 ( \ln \frac{m_c^2}{\Lambda_{\chi}^2} + \frac{5}{3})  \; .
\label{gQ4S}
\ee  
Using the same method as for the $Q_6$, we obtain for a variable
penguin coefficient within dimensional regularization (Only Fig.3a 
contributes in this case, but with axial currents acting at the weak
 vertices):
\be
g(Q_4)_{DR} =   \frac{\alpha_s}{12 \pi} \,
\left( \ln \frac{m_c^2}{M^2} + \frac{2}{3}  
\, - 9 \frac{M^2}{\Lambda_{\chi}^2}
 \left[\ln \frac{m_c^2}{M^2} \, \ln \frac{m_c^2}{\Lambda_\chi^2}
+ \Delta  \right] \right)  \; ,
\label{gQ4DR}
\ee  
where $\Delta$ parametrizes some (calculable) non-leading terms, and we
have put $\mu = \tilde{\mu} = \Lambda_\chi$ in (\ref{drpeng}).
 Using the values $m_c =$ 1.4 GeV, 
$M \simeq$ 200 MeV and $\Lambda_\chi \simeq$ 830 MeV,
 we obtain an  increase by almost 70$\%$ of the leading term with
 respect to (\ref{gQ4S}).
However, this is partially compensated by the non-leading term 
$\sim M^2/\Lambda_\chi^2$, and the overall result is an
 increase of order 20 $\%$ of $g(Q_4)_{DR}$ with respect
 to $g(Q_4)_S$. Even if
$g(Q_4)$ is small ($\simeq$ 0.05), (\ref{gQ4DR}) gives an idea of how
the penguin contributions behave when the  variation of the penguin
coefficients are taken into account. However,  $g(Q_4)_{DR}$ is rather
 sensitive to variations in $M$.

\vspace{0.3cm}

Both for $Q_6$ and $Q_4$, our procedure makes the $\chi QM$
calculation sensitive to $m_c$, which does not sound unreasonable because
 the scale $m_c$ is not too far above $\Lambda_\chi$.
However,  even if matrix elements of the penguin contribution
 are sensitive to  the scale $m_c$,  
 we expect that  the effect of 
 high momenta $\sim m_c$ should be
 damped. Especially, we expect
the leading behaviour $\sim m_c^2$ in (\ref{gQ6BE}) to be damped
 due to some cut-off $\Lambda$.
Therefore we have studied the $Q_6$ amplitude by using proper time 
regularization. 
 Then the two loop integral can not be solved analytically, but we
can perform a  numerical calculation for various values of the
cut-off $\Lambda$.
This way of regularization will have an  exponential damping of
higher momenta. For this calculation we have used (\ref{fpeng})
with $\alpha_s = \alpha_s(\Lambda)$.

\begin{table}
\begin{center}
\begin{tabular}{|c|c|c|c|c||c|c||c|}  \hline
$\Lambda$& $M$ &$\alpha_s(\Lambda)$&$f_\pi$
&$ \langle \bar{q}q \rangle^{1/3}$ & $g(Q_6)_P^0$ & $g(Q_6)_P$ 
 & $g(Q_6)_P^M$  \\ \hline
700 & 330 & 1.00 & 101 & $-189$ & -0.008  & -0.006 & - 0.01     \\ \hline
830 & 330 & 0.70& 126 &   $-222$ &0.03 & 0.04 &0.03    \\ \hline
1000 & 330 & 0.53 & 155 & $-261$ &0.10 & 0.11 &0.05   \\ \hline
1200 & 330 & 0.44 & 185 & $-303$ &0.19 & 0.22 &0.06   \\ \hline
1400 & 330 & 0.38 & 211 & $-342$ &0.30 & 0.36 &0.07    \\ \hline  \hline
700 & 250 &1.00  &  82 &  $-185$ &0.04 & 0.05 &0.09  \\ \hline
830 & 250 &0.70  &  98 &  $-213$ &0.08 & 0.09 &0.11    \\ \hline
1000 & 250 & 0.53 & 115 & $-248$ &0.15 & 0.16 &0.12   \\ \hline
1200 & 250 & 0.44 & 133 & $-285$ &0.23 & 0.26 &0.13   \\ \hline
1400 & 250 & 0.38 & 148 & $-320$ &0.30 & 0.37 &0.12   \\ \hline     \hline
700 & 150 &1.00  &  47 &  $-167$ &0.09 & 0.10 &0.42   \\ \hline
830 & 150 &0.70 &  53 &   $-190$ &0.11 & 0.12 &0.37     \\ \hline
1000 & 150 & 0.53 &  60 & $-218$ &0.14 & 0.16 &0.33  \\ \hline
1200 & 150 & 0.44 &  66 & $-248$ &0.18 & 0.22 &0.31   \\ \hline
1400 & 150 & 0.38 &  72 & $-276$ &0.21 & 0.27 &0.28  \\ \hline
\end{tabular}
\caption{Values of  $g(Q_6)_P^0$ and $g(Q_6)_P$, for constant and
momentum dependent penguin coefficient respectively, calculated
for different values of $\Lambda$ and $M$. The corresponding modified 
values  $g(Q_6)_P^M$obtained when compensating for wrong values obtained for
$f_\pi$ and $<\bar{q} q>$ are also given.} 
\label{constants}
\end{center}
\end{table}

\vspace{0.2cm}

The values found for
 $g^{(1/2)}_{8}(Q_6)_{P} \equiv g(Q_6)_{P}$ for finite cut-offs
 $\Lambda$ calculated within proper time regularization are tabulated in
table 1. For comparison, the corresponding values $g^{(1/2)}_{8}(Q_6)_P^0
 \equiv g(Q_6)_P^0$ for a constant penguin coefficient (using (\ref{nlpeng})
 with $\mu = \Lambda$) are also given.
As mentioned above, the integral for $g(Q_6)$ contains the product
 of a quadratic divergence corresponding to the quark
 condensate (see (\ref{qcond}))
and  a logarithmic divergence for $f_\pi$ (see (\ref{fpi})).
One should note that for various values of $\Lambda$ and $M$, the 
obtained values for   $f_\pi$ and $<\ov{q} q>$ are rather
different from the physical values  (The gluon condensates
in (\ref{fpi}) and (\ref{qcond}) are not included here).
This leads to wrong values for
 $g(Q_6)$ both for a constant and a momentum dependent penguin coefficient.
To compensate for this, we have calculated the modified values
 $g(Q_6)_{P}^M$ obtained when multiplying with the physical
$f_\pi$ and  $<\ov{q} q>$, and dividing with the respective values 
$f_\pi(\Lambda,M)$ and   $<\ov{q} q>(\Lambda,M)$ in table~1.
This procedure should correspond to the necessity of keeping  the relation
$f_\pi^{(0)} = f$ as mentioned below (\ref{rho}).
 For the
physical quark condensate, we have used $<\ov{q} q>(\mu=\Lambda)$ 
from formula (\ref{BBGcond}), with 
$<\ov{q} q>(\mu = 1 {\mbox GeV}) \, \simeq  (-235 {\mbox MeV})^3$, 
corresponding to $(m_u + m_d) \simeq$ 12 MeV at $\mu =$ 1 GeV.

\vspace{1cm}

\begin{Large}

6. Discussion

\end{Large}

\vspace{0.2cm}

 We have  investigated the consequences of taking into account the momentum
dependence of the penguin coefficient for  $K \rightarrow 2 \pi$
decays.
The calculations are performed within the Chiral Quark Model ($\chi QM$).
It might seem surprising that the  penguin amplitude corresponding to
$Q_6$ gives such a big
leading contribution within dimensional regularization.
But we
 will   point out that a  similar result was
 obtained\cite{siamp} for the 
so called siamese penguin diagram, where we found
 a result $\sim m_c^2$ in the limit $\Lambda^2 \rightarrow \infty$,
instead of the ``expected'' $\sim \Lambda^2$. 
Unfortunately the dimensional regularization case, which is 
the only one which can be done
analytically for $Q_6$,  is hard to analyze numerically because dimensional 
regularization does not distuinguish quadratic and logarithmic divergences.
One might also worry if dimensional regularization includes too high
loop momenta within the $\chi QM$. On the other hand, the dimensional 
regularization result for $Q_4$ seems to be reasonable.
 
For finite values of the cut off we have performed a numerical integration. 
 Using proper time regularization, we obtain a $10-20 \%$ increase
of the penguin contribution compared to the case where the penguin
 coefficient is considered as momentum independent.
 The reason that we get a slightly bigger
value for $g(Q_6)_P$ than  $g(Q_6)_P^0$ is the following: The relevant
diagrams for the $K \ra \pi$ transition to order $k^2$ are the "eight"
diagram (Fig.3a) and the "keyhole" diagram (Fig.3b). For a
constant penguin coefficient, it is a
partial cancellation between these diagrams (in the order $k^2$ terms)
which makes the small $L_5$ in (\ref{cL5}). For a momentum {\em dependent}
penguin coefficient, the "eight" and the "keyhole" diagrams
are influenced in different ways, and $g(Q_6)_P$ is not
anymore  directly proportional to $L_5$, but to some
 "smeared out" and slightly bigger analogue of $L_5$.

We have not considered the effect of the momentum dependence of the
coefficients $C_{1,2}$ (or equivalently $C_{\pm}$) of the non-penguin 
operators $Q_{1,2} \, (Q_{\pm})$ in (\ref{QLL}). However, these coefficients
are not so sensitive to the involved parameters, and we expect even smaller
effect of momentum dependence than for penguin-operators.
Furthermore, we have not taken advantage of the renormalization
group analysis involving the penguin coefficient $C_6$. This is because 
 the individual quark momenta can hardly be
distinguished, owing to the mixing of operators.

Our analysis offers no solution to the $\Delta I=1/2$ puzzle
for $K \ra 2\pi$ decays. Moreover, the 10-20$\%$ increase of the
penguin contribution seen in table 1 is moderate compared to the
increase in the penguin coefficient $C_6$ owing to two loop
 effects \cite{BuCi}.
The penguin contribution to $K \ra 2\pi$ is very sensitive to the
involved parameters as the quark condensate, the constituent
quark mass $M$, and the coefficient $C_6$ through $\Lambda_{QCD}$
and the matching scale $\mu$.
In conclusion, other contributions than $g^{(1/2)}_{8}(Q_6)$
 must be important for  the  $\Delta I=1/2$ enhancement \cite{Trst}.

\vspace{1cm}

\begin{Large}

7. Appendix

\end{Large}

\vspace{0.2cm}

We will here give
some details from the calculation of $I_{\chi}$ in (\ref{chint}).
 We expand the propagators up to second order in the
external virtual meson momentum $k$. Then the propagators $[r^2-M^2]^{-1}$
and $[(r-p)^2-M^2]^{-1}$ occur in some powers. We need one Feynman parameter
$(x)$ to perform the integral. We use the general formula
\begin{equation}
 \int \frac{d^D r}{(2\pi)^D}
 \, [r^2 - A]^{-n} \; = \; i J_n(D) [-A]^{(D/2-n)} \; ,
\label{appI1}
\end{equation}
where
\be
J_n(D) \; = \; \frac{(-\pi)^{D/2}}{(2\pi)^D} \,
\frac{\Gamma(n-D/2)}{\Gamma(n)} \; .
\ee
Note that $J_{n+1}(D) = J_n(D) (n-D/2)/n$. Dropping the constant term
$\sim k^0$, we obtain

\begin{eqnarray}
I_{\chi} & = &   
i J_3(D) k^2 \int_0^1 dx \, \left\{ S_c(3) \left[ \frac{6}{D}(1-x(1-x)) -
2 \right. \right. \nonumber \\ & & \left. \left. - \; (\frac{6-D}{D})(1-2x(1-x)) \right] \, 
+  (\frac{6-D}{D}) x(1-x) \, M^2 \, S_c(4) \,  \right\} \nonumber \\ & &
 - \; (c~\ra~u) \; ,
\label{appI2}
\end{eqnarray}
where the quantity $S_q(n)$ depends on the current quark masses for
$q=c,u$:
\be
S_q(n) \; = \; 
\int \frac{d^D p}{(2\pi)^D} \,  [C_P(p^2)]_q \, [-A]^{(D/2-n)} \;
,
\label{appI3}
\ee
where $[C_P(p^2)] =  [C_P(p^2)]_u -  [C_P(p^2)]_c$,
and $A = M^2~-~x~(1~-~x)~p^2$. To find $S_q(n)$, we
have to use the dimensional regularization version of  $[C_P(p^2)]$
in (\ref{drpeng}), and we need the generalized Feynman parametrization
\be
\frac{1}{(-A)^l (-D_q)^{\epsilon}} \;= \; \frac{1}{B(l,\epsilon)}
 \, \int_0^1 dy \, \frac{(1-y)^{l-1} y^{\epsilon-1}}{N_q^{l+\epsilon}}
\label{GFeyn}
\ee
where $l= n-D/2$ and $N_q = (1-y) (-A) + y (-D_q) = h (p^2 - \ov{m_q}^2)$.
Here, $h = y t (1-t) + (1-y) x(1-x)$ and 
$\ov{m_q}^2 = (y m_q^2 + (1-y) M^2)/h$. The quantity $B$ is given by
\be
B(a,b) \; = \; \int_0^1 \, dx \, x^{a-1} \,  (1-x)^{b-1} \; = \; 
\frac{\Gamma(a) \cdot \Gamma(b)}{\Gamma(a+b)}
\ee
Using again (\ref{appI1}), we obtain
 \begin{equation}
S_q(n) \; = \;  i J_{l+\epsilon}(D) \frac{\Gamma(\epsilon)}{B(l,\epsilon)}
\int_0^1 dt \, 6 t (1-t) \int_0^1 \, dy \, 
\frac{(1-y)^{l-1} y^{\epsilon-1}}{h^{l + \epsilon}} \, 
(-\ov{m_q}^2)^u \; ,
\label{Sint}
 \end{equation}
where $u = D/2 - (l+\epsilon)$. Now, for the charm quark case $q=c$,
we use the approximation $\ov{m_c}^2 = y m_c^2 /h$. Then we obtain
 \begin{equation}
S_c(n) \; = \;  i J_{l+\epsilon}(D) \frac{\Gamma(\epsilon)}{B(l,\epsilon)}
(-m_c^2)^u \int_0^1 dt \, 6 t (1-t) \, I(x,t)
\label{Sint1}
 \end{equation}
 Then the clue
is to recognize the integral
\begin{equation}
I(x,t) \; = \;  
\int_0^1 \, dy \, 
\frac{(1-y)^{l-1} y^{\epsilon + u -1}}{h^{u + l + \epsilon}} \; =
\; \frac{B(l,u+\epsilon)}{[x(1-x)]^l [t(1-t)]^{u+\epsilon}}
\label{Sint2}
 \end{equation}
which gives
\begin{equation}
S_c(n) \; = \;  \ov{S_c(n)} \, [x(1-x)]^{-l} \;
\label{Sint3}
\end{equation}
with
\begin{equation}
\ov{S_c(n)} \; = \;  i J_{l+\epsilon}(D) \frac{\Gamma(\epsilon)}{B(l,\epsilon)}
(-m_c^2)^u \, B(l,u+\epsilon) \, 6 \, B(2-u-\epsilon,2-u-\epsilon)
\label{Sint4}
 \end{equation}
Then, to leading order in $m_c^2$, we find
\begin{equation}
I_{\chi} \; = \;  
 - i J_3(D) k^2 \ov{S_c(3)} \, B(D/2-2,D/2-2) \; 
 \end{equation}
Here the factor $B(D/2-2,D/2-2)$ corresponds to the logarithmic
divergence connected to $f_{\pi}$, and $J_{l+\epsilon}(D)$
in $S_c(3)$ to the quadratic divergence connected to the quark condensate.



\newpage

\begin{figure}
\begin{center}
\mbox{\epsfig{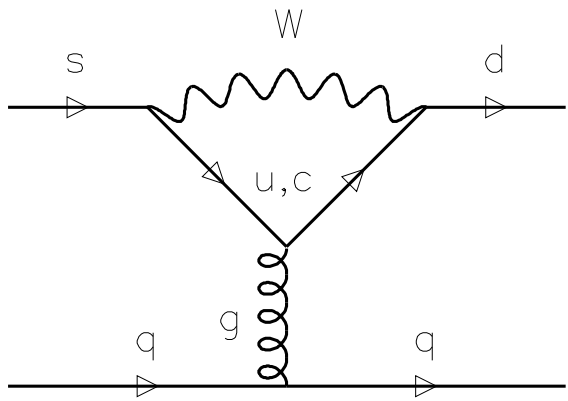}}
\caption{The penguin loop diagram. At the lower line $q=u,d,s.$}
\end{center}
\end{figure}

\begin{figure}
\centerline{\psfig{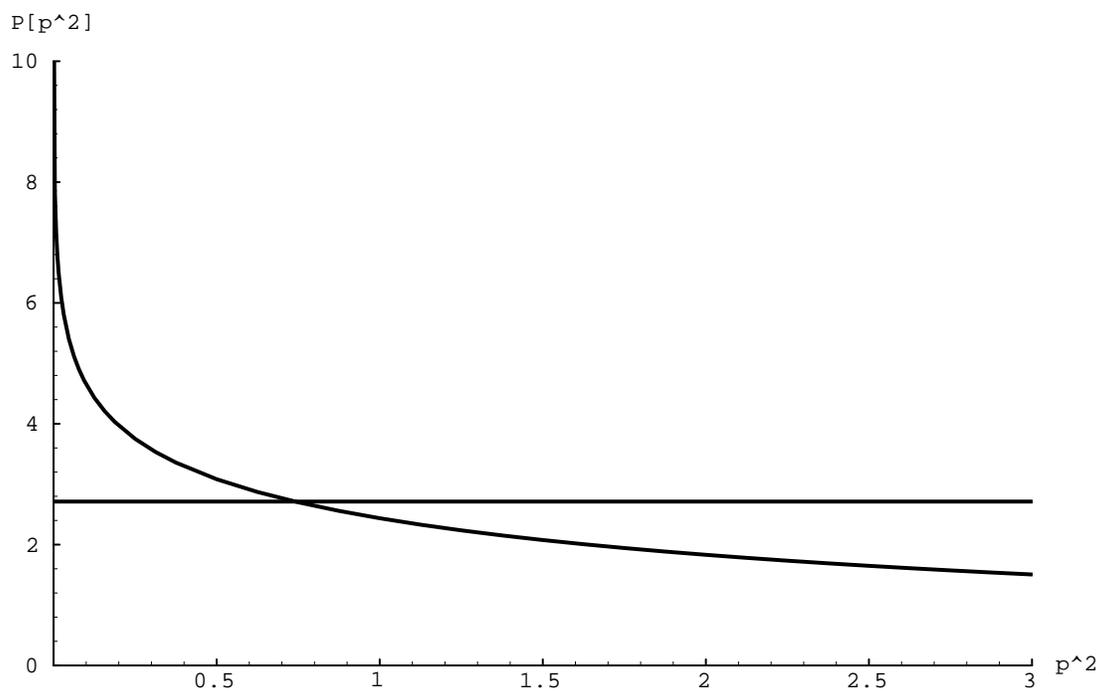}}
\caption[]{Variation of the penguin coefficient 
$C_P = - \frac{\alpha_s}{3 \pi} \, P(p^2)$  in
(6) with squared Euclidean momentum $p^2$ (-in units $\mbox{GeV}^2$).
The horisontal line represents the value in eq. (\ref{nlpeng})
for $\mu=$ 0.83 GeV. }
\end{figure}

\begin{figure}
\begin{center}
\mbox{\epsfig{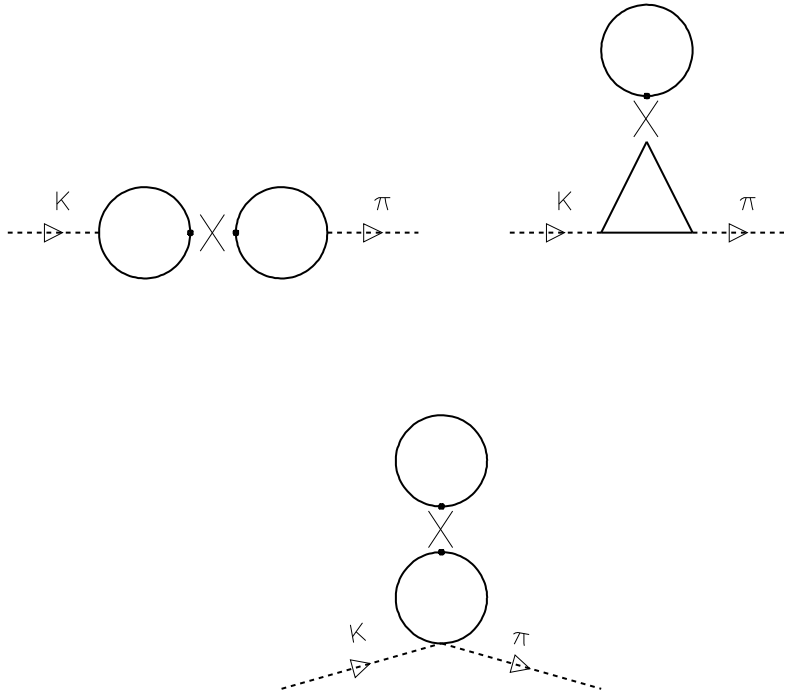}}
\caption{Diagrams for the penguin operator $Q_6$ contribution to
$K \ra \pi$. The solid lines represent quarks.Diagram (a), upper left,
 is the ``eight'' diagram, while 
(b) is the ``keyhole'' diagram. Diagram (c) serves to cancel the momentum 
independent part of (a) and (b). }
\end{center}
\end{figure}

\end{document}